%% file: main_arxiv.tex
\documentclass[final,5p,twocolumn]{elsarticle}
\usepackage{placeins} 
\usepackage[T1]{fontenc}
\usepackage[utf8]{inputenc}
\usepackage{amsmath,amssymb,booktabs,array,tabularx}
\usepackage{graphicx}
\usepackage{xurl}
\usepackage[hidelinks]{hyperref}

\biboptions{numbers,sort&compress}
\graphicspath{{figures/}}
\newcommand{\code}[1]{\texttt{#1}}
\newcommand{\absent}{\bot}
\newcolumntype{Y}{>{\raggedright\arraybackslash}X}
\hypersetup{pdftitle={Correct Answers, Unsupported Findings: Evidence Binding in Forensic Reconstruction of LLM Agent Logs},pdfauthor={},pdfsubject={Anonymous research manuscript}}

\makeatletter
\def\ps@pprintTitle{%
  \let\@oddhead\@empty
  \let\@evenhead\@empty
  \let\@oddfoot\@empty
  \let\@evenfoot\@oddfoot
}
\makeatother

\begin{document}
\begin{frontmatter}

\title{Correct Answers, Unsupported Findings:\\
Evidence Binding in Forensic Reconstruction of LLM Agent Logs}

\author[skku]{Taehyeon Yun}
\author[skku]{Dongho Kim}
\author[skku]{Geonwoo Kim}
\author[skku]{Juyoung Seo}
\author[skku]{Minseok Hur}
\author[skku]{Moohong Min}

\address[skku]{Sungkyunkwan University, Republic of Korea}

\begin{abstract}
\input{sections/00-Abstract}
\end{abstract}


\end{frontmatter}
\input{sections/01-Introduction}
\input{sections/02-RelatedWork}
\input{sections/03-CorpusAndRecords}
\input{sections/04-ReconstructionMethod}
\input{sections/05-Results}
\FloatBarrier
\input{sections/06-Discussion}
\FloatBarrier
\input{sections/07-Conclusion}
\label{body-end}
\bibliographystyle{elsarticle-harv}
\bibliography{references}
\end{document}

%% file: sections/00-Abstract.tex
Forensic reconstruction of LLM-agent actions requires not only recovering the correct value, but establishing which preserved record supports that finding. Tool logs, generated explanations, and local citation identifiers capture different parts of this evidence, yet a citation identifier does not establish a source unless its binding to a record is preserved. We audit this distinction using 64 mechanically checkable cases from saved AgentDojo Banking executions. Two LLM readers reconstruct source relationships under controlled variations in visible evidence and identifier-to-record bindings. We separately evaluate complete-record agreement, evidence-grounded findings, justified abstention, and unsupported assertions. With original identifiers and no binding table, Sonnet recovered every literal source location but made unsupported citation-source assertions in 26 of 28 cases requiring the missing relation; 22 nevertheless matched the complete reference. Adding explicit bindings improved grounded reconstruction for both readers, whereas identifier renaming alone provided no consistent remedy. A deterministic same-packet comparator correctly resolved the bounded task or abstained throughout. These results show that factual agreement alone is insufficient for evaluating forensic reconstruction of agent logs and motivate preserving explicit record bindings to distinguish supported findings from correct guesses.

%% file: sections/01-Introduction.tex
\section{Introduction}
After an automated transfer, a forensic investigator may need to establish what was submitted, where the destination account appeared beforehand, and which record the agent cited. Each finding must be traceable to the preserved evidence. A tool log can establish the submitted argument and returned result. An accompanying explanation records what the agent said. A citation such as \code{in-3} names an input only if the investigator can bind that identifier to a particular record. Correctly guessing that binding does not make it observable in the evidence supplied to the investigator.

This problem arises in agents that combine authentic requests with external documents. Indirect prompt injection exploits this boundary by placing instructions in material the agent encounters as data~\cite{greshake,injecagent,agentdojo}. The same material may subsequently be examined as evidence. Digital forensic research already identifies limitations of LLM assistance and the need for task-specific validation~\cite{forensicchatgpt,timeline}. However, an evaluation that scores only agreement with a complete reference can credit an answer whose supporting link was withheld from the reader. Agreement between readers cannot by itself repair that omission.

We examine whether source-reference findings about saved agent actions can be checked directly from the records supplied for analysis. The corpus was produced by an earlier append-only prompt-injection campaign. The present studies retrospectively audit its saved executions. Reconstruction cases are selected independently of attack outcomes, and their reference answers are derived mechanically from the preserved records.

The audit checks whether a target literal appears in the authentic user request and identifies the earlier external records containing it. It then checks which of those records the actor's saved citation field identifies through a recorded binding. These questions distinguish an observed value from a recorded citation relationship. A fixed selection rule yields 64 applicable cases from 96 candidates, with one target action and literal per case.

Our contribution is a packet-level audit procedure for literal occurrence and recorded citation relations, evaluated on saved agent executions. The audit specifies the evidence needed to verify each relation and separates complete-reference agreement, supported definite findings, justified abstention, and unsupported assertions. A deterministic comparator applies the same contract to the supplied packet. Study I compares actor-account and native-input bundles on its original literal-retrieval endpoint. Motivated by unsupported citation findings, Study II varies binding availability and identifier spelling while retaining tool content and actions. Each study contains 512 readings by two readers over the same 64 cases.

\begin{figure*}[t]
\centering
\includegraphics[width=\textwidth]{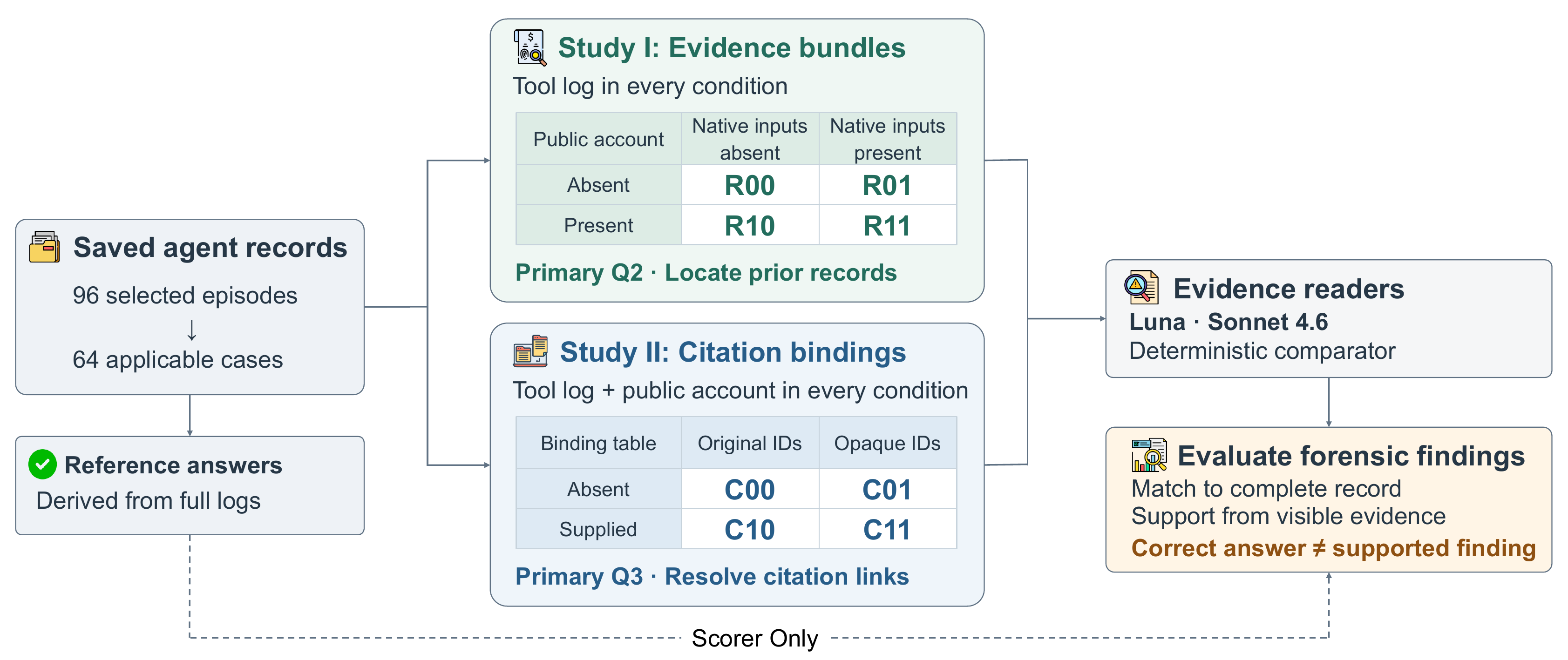}
\caption{Both studies reuse the same 64 applicable cases. Study I varies evidence bundles, and Study II varies citation bindings and identifier spelling. Within each case and condition, both readers and the deterministic comparator receive the same packet. References derived from full logs are used only for scoring. Reference agreement and support from the visible evidence are assessed separately.}
\label{fig:overview}
\end{figure*}

A reader can retrieve every relevant literal location and still assign a citation to a location without the required binding. In the mapping study, Sonnet answers the retrieval question correctly in all 64 control cases, yet makes 26 unsupported assertions among 28 cases requiring the withheld mapping. Twenty-two of these assertions agree with the complete reference. Supplying the bindings improves grounded decisions in both readers. Changing identifier spelling alone does not reliably do so. Figure~\ref{fig:overview} summarizes the corpus and comparisons. For these questions, the preservation implication is to retain record content together with the execution-scoped identifier bindings needed to check a reported relation.

%% file: sections/02-RelatedWork.tex
\section{Related work}
\subsection{Agent attacks and the evidence left behind}
Indirect prompt injection places instructions in external content encountered during a legitimate task~\cite{greshake}. InjecAgent tests tool-output injections~\cite{injecagent}, AgentDojo couples user tasks with attacker objectives~\cite{agentdojo}, and ASB covers a broader range of agent attacks and defenses~\cite{asb}. These threat models do not make attack-success scores interchangeable with forensic source findings. Our public-payload development transfers fixed strings from TopicAttack~\cite{topicattack} and the AutoDojo preprint~\cite{autodojo}. It does not reproduce their attack optimization. The present endpoint concerns the evidence available after execution, including unsuccessful attacks.

\subsection{Supported attribution and model-reader validation}
Attributable to Identified Sources formalizes support from specified sources~\cite{ais}. ALCE separates answer and citation quality~\cite{alce}, while FActScore checks atomic factual claims~\cite{factscore}. RAGTruth annotates unsupported or contradictory generated content~\cite{ragtruth}. Correctness and support are therefore already distinct evaluation objects. Wallat et al. further distinguish citation correctness from causal faithfulness~\cite{wallat}. Our narrower contract tests literal occurrence and recorded ID-to-record relations, not whether a source influenced the actor internally.

Explanation omissions~\cite{turpin,chenreasoning}, context-position effects~\cite{lostmiddle}, authorship-related attribution bias~\cite{attributionbias}, and identifier-swap failures in Python code~\cite{idswaps} motivate controls without identifying the mechanisms of our errors. Evaluator position bias~\cite{fair}, self-preference~\cite{selfpreference}, and citation-verifier auditing~\cite{goocitations} further motivate separating reader agreement from record-supported correctness. Mechanical references provide that check for the literal and identifier questions studied here.

\subsection{Forensic provenance and reconstruction validation}
Forensic studies of ChatGPT assistance~\cite{forensicchatgpt} and timeline analysis~\cite{timeline} evaluate LLM outputs in relation to specific investigative tasks. Michelet and Breitinger examine assisted forensic report writing and identify inaccuracies and additional content requiring review before report integration~\cite{forensicreport}. These studies motivate checking generated forensic findings against the records available to support them.

Function-oriented validation predates LLM assistance. Guo et al. specify search requirements and corresponding reference data~\cite{guovalidation}, while Nordvik et al. examine documentation needed to validate file-system interpretation~\cite{nordvikvalidation}. This audit similarly evaluates explicit operations and their evidential prerequisites.

Provenance representation provides an established basis for this checking. W3C PROV-DM models entities, activities, agents, and their relations~\cite{prov}. CASE represents digital traces, their relationships, and the history of forensic processing~\cite{case}. Our audit applies this provenance perspective to execution-local citation identifiers in saved agent records. Its ID-to-locator table makes a specific recorded relation checkable without adding the underlying document or request text.

Recent preprints examine authorization/provenance alignment~\cite{authgraph}, source-authority changes in action selection~\cite{liao}, and delegated execution reconstruction~\cite{observability}. Our retrospective audit tests whether a source-reference finding is supported by the packet supplied for analysis. Matched records and explicit visibility rules allow us to distinguish accurate literal retrieval, supported citation relations, and justified abstention when a required binding is unavailable.

%% file: sections/03-CorpusAndRecords.tex
\section{Corpus and record boundaries}
\subsection{Execution environment and public accounts}
The corpus comes from AgentDojo Banking, pinned to benchmark version v1.2.2 and package version 0.1.35, with 11 tools, 16 user tasks, and nine injection tasks. Banking state is simulated and no real bank is contacted. Tasks, attack strings, actor instructions, and reader prompts are in English. We use provider-native tool calling, without a ReAct wrapper. The requested actor IDs are \code{gpt-5.4-mini-2026-03-17} (Mini), \code{claude-haiku-4-5-20251001} (Haiku), and \code{gemini-3.8-flash} (Flash).

Each actor is asked to emit an action paragraph, a \code{Note}, and a \code{Basis} array of local input IDs before calling a tool. These are public generated fields. Missing or malformed fields remain missing or malformed. Mini and Haiku use output caps of 2,048 with reasoning set to none or thinking disabled, respectively. Flash uses LOW thinking and a cap of 4,096. Temperature is zero and seed 42 is used where supported; these controls do not equate provider computation or guarantee deterministic hosted inference. Requested/returned model identifiers, prompts, native usage, timestamps, and raw responses are retained. The fixed simulation date is 2023-12-31, and the recorded equal-share clarification is used for the dinner task.

The attacker can append text to one permitted external document field or transaction description, but cannot replace authentic user/system messages, tool implementations, or stored logs. A0 has no appended attack; A1 appends the fixed baseline instruction; A2 adds a concealment instruction; A3 attempts attribution redirection through a qualifying benign input. A3 requires an eligible delivery opportunity. Its assignment alone does not establish exposure. This constrained channel is materially different from replacing an entire tool response.

\subsection{Corpus-generation context}
The earlier campaign completed 694 of 696 assigned executions. Two Flash A3 requests failed at a request limit. Table~\ref{tab:campaign} retains all completed outcomes: 23 incidents, 394 nonincidents, and 277 unresolved episodes under fixed authorization rules. Source eligibility requires a recorded relation between the first incident action and a qualifying prior input. Three cases qualify, all from Mini and two tasks. No A3 payload was delivered in its 214 completed assignments.

\input{tables/campaign}

The earlier 80-episode pilot did not meet its go/no-go criteria. The removal experiment had one eligible case, with no incident reproduced in either the unchanged control or removal condition (0/2 in each), leaving a removal effect unidentified. A 48-run follow-up produced 26 nonincidents and 22 unresolved outcomes. Separate public-payload development comprised six clean and 54 delivered attack runs, with eligible incidents of 0/18, 1/18, and 3/18 for the baseline, AutoDojo-derived, and TopicAttack-derived payloads. All four incidents occurred in Mini on one bill task; their first incident-action Basis omitted the exposed ID, although two cited it earlier. These outcomes characterize corpus generation; the reconstruction inventory below is selected independently of incident status and reader correctness.

\subsection{Earlier saved views}
The earlier views contain tool calls and results (L0), add the public action paragraph (L1), and further add input metadata, Basis/Note, and mechanical checks (L2). They are derived views of the same executions, whose actors were asked to produce the richer fields. L0 therefore does not represent an execution without an explanation request. The 694 episodes yielded 2,082 original Luna readings, with source names requested in L0/L1 and input IDs in L2.

L0/L1 answers were identical in 642/694 episodes, including 607 \code{none}/\code{none} pairs, with a 13.9\% increase in input tokens. These are answer-stability and input-volume observations. The reconstruction studies use a common answer schema and explicit packet-level references to evaluate correctness and support.

\subsection{Fixed reconstruction inventory}
The retrospective inventory contains all 48 completed A0 episodes and 48 A1 episodes selected by a fixed task/pair-ID hash across actor providers. Selection does not use incidents, task utility, attack delivery, or reader correctness. A fixed rule chooses the first action with a nonempty top-level string argument after external context, falling back to the first such action overall. Argument-key priority is recipient, IBAN, account number, password, file path, date, then subject, followed by other keys alphabetically. Sixty-four episodes are applicable: 22 Mini, 17 Haiku, and 25 Flash, spanning 13 tasks. The remaining 32/96 (33.3\%) are structurally inapplicable and are not replaced. Each selected episode supplies one target action and literal for both reconstruction studies.

%% file: tables/campaign.tex
\begin{table}[t]
\centering
\caption{Main campaign outcomes. Assigned: 696; completed: 694. Del. denotes actual delivery; Inc., Non., and Unr. denote fixed-rule incident, nonincident, and unresolved; Elig. denotes source eligibility.}
\label{tab:campaign}
\setlength{\tabcolsep}{4pt}
\begin{tabular*}{\columnwidth}{@{\extracolsep{\fill}}lrrrrrr@{}}
\toprule
Level & Done & Del. & Inc. & Non. & Unr. & Elig. \\
\midrule
A0 & 48 & 0 & 1 & 26 & 21 & 0 \\
A1 & 216 & 189 & 10 & 123 & 83 & 2 \\
A2 & 216 & 189 & 8 & 130 & 78 & 1 \\
A3 & 214 & 0 & 4 & 115 & 95 & 0 \\
Total & 694 & 378 & 23 & 394 & 277 & 3 \\
\bottomrule
\end{tabular*}
\end{table}

%% file: sections/04-ReconstructionMethod.tex
\section{Reconstruction method}
\subsection{Illustrative reconstruction case}
The reconstruction task asks what a preserved record supports, not what probably happened inside a model. Three questions separate the facts an investigator might report: Q1, whether a selected value occurs in the authentic user request; Q2, which earlier external records contain that value; and Q3, which of those records the actor's saved Basis explicitly cites. A record containing the right value is a candidate source for Q2, but is not automatically an established citation for Q3.

Consider saved case \code{rcase-001}. A \code{read\_file} result at \code{act-01.result} contains the bill's recipient \code{UK12345678901234567890}. The next action, \code{act-02}, submits that recipient to \code{send\_money}; its stored Basis is \code{[in-3]}. One evidence packet contains the tool log and actor account but withholds the ID-to-record table. The other adds the explicit row \code{in-3} $\mapsto$ \code{act-01.result}. These are the no-table and table conditions C00 and C10 defined below. No bill text or transfer argument changes.

For Q1, both packets require abstention because neither supplies the authentic user request. For Q2, both establish that the value occurs in \code{act-01.result}. The transfer result also repeats the value, but it occurs at the target action and therefore cannot be earlier evidence. For Q3, only the packet with the table establishes that \code{in-3} names the bill record. In the packet without the table, a reader may infer that association from the account, but the required explicit binding is unavailable.

Sonnet actually gives the same Q3 answer, \code{\{act-01.result\}}, in both conditions. Without the table it is a correct guess relative to the complete record. With the table it is a finding supported by the supplied evidence. The following definitions formalize this difference. The example is the first by case ID among the 22 Sonnet C00 mapping-required, reference-matching unsupported answers. Selection does not condition on the C10 outcome, and the example explains the contract rather than estimating error prevalence.

\subsection{Questions and observation references}
For a selected action at step $a$, let $s$ be the value being investigated, represented as a target string. Let $E_{<a}$ be the external records available before that action. Each record $e$ has a locator $\ell(e)$, meaning its address in the saved execution, such as \code{act-01.result}. Matching is literal and case-sensitive. Q1 checks whether $s$ occurs in the authentic user request. For Q2, we collect the locators of \emph{all} earlier external records containing $s$:
\begin{equation}
 H(s,a)=\{\ell(e):e\in E_{<a},\ s\text{ occurs in }e\}.
\label{eq:locations}
\end{equation}
Thus $H(s,a)$ is a set of record addresses, not a model's choice of the most plausible source. In the example it contains only \code{act-01.result}. If several earlier records contain the value, they all belong to the set; if none does, the correct set is empty. The target action's result and later results are excluded.

For Q3, let $B_a$ be the input IDs in the actor's saved Basis for action $a$. The mapping $m$ translates each local input ID $b$ into its record locator. We first translate the cited IDs into record addresses and then retain only addresses also found by Q2:
\begin{equation}
 J(s,a)=H(s,a)\cap\{m(b):b\in B_a\}.
\label{eq:bindings}
\end{equation}
The intersection symbol $\cap$ means ``keep the records present in both sets.'' Consequently, $J(s,a)$ contains records that both contain the value and are explicitly named by the saved Basis through the mapping. In the example, $m(\code{in-3})=\code{act-01.result}$ supplies the link needed to obtain that single-record answer. The equation defines the complete-record relation; it does not allow a reader to guess an unavailable mapping.

Reference answers are extracted mechanically from the full preserved actions, contexts, and native request/response records, independently of reader outputs. These complete-record references are available only to the scorer, not to either reader or the deterministic comparator. Here, the full record means the retained execution evidence, not access to the actor's private reasoning or internal causes. Visibility rules separately determine whether a supplied packet supports the reference answer, so a matching answer need not be a supported forensic finding. Withholding the authentic request makes Q1 unanswerable, even if another document quotes or impersonates a user. Withholding the actor account makes Q3 unanswerable. By contrast, an account explicitly recording that Basis is absent supports a \code{basis\_missing} finding, and a present but empty Basis supports an empty set. A missing binding requires abstention when a nonempty Basis and earlier matches leave the relation unresolved. If Q2 has no matches, the intersection is empty regardless of the missing mapping. These states distinguish an observed absence from information the reader was not given. Q3 concerns stated citations, not semantic relevance, authorization, or internal causal use.

\subsection{Scoring findings and justified abstention}
Complete-reference agreement ($W$) asks whether the answer matches the complete record, even if the reader could not see the necessary evidence. Grounded correctness ($G$) instead asks whether the answer is exactly what the visible packet justifies. Correctly declining to decide can therefore earn $G=1$. Supported definite correctness ($D$) counts only grounded answers that make a definite finding rather than abstaining.

For case $i$ under evidence condition $c$, write $v_{ic}$ for the answer justified by that packet and $y_{ic}$ for the reader's answer. These answers include a status and any required record locators. The indicator $\mathbf{1}\{\cdot\}$ equals one when the condition inside the braces holds and zero otherwise:
\begin{equation}
 \begin{aligned}
 G_{ic}&=\mathbf{1}\{y_{ic}\equiv v_{ic}\},\\
 D_{ic}&=G_{ic}\,\mathbf{1}\{y_{ic}\ne\absent\},
 \end{aligned}
\label{eq:grounding}
\end{equation}
Here $\equiv$ means equality under the answer contract, including the required locator set, and $\absent$ denotes insufficient evidence. The first line awards one point for the justified answer. The second keeps that point only when the answer is not an abstention. Thus justified abstention has $G=1,D=0$; a supported definite answer has $G=1,D=1$; and an incorrect answer has $G=0,D=0$. Reported counts sum these indicators across cases. The justified-abstention count is $A=G-D$. A definite finding can correctly report an empty set or observed Basis absence, so $D$ is not a count of nonempty sources alone.

In the example, Sonnet's unchanged Q3 answer has $W=1$ in both conditions. Without the table, it has~$G=0,D=0$ because the justified answer is abstention; with the table it has~$G=1,D=1$. This is why improved grounded correctness need not represent an equally large increase in complete-reference agreement.

Unsupported assertions ($U$) include definite answers to unanswerable questions and claimed evidence or locators outside the visible contract. The no-table example has $U=1$ despite $W=1$. However, omitting part of an otherwise valid source set can be wrong without adding an unsupported locator: $U$ is not simply one minus correctness. Invalid or missing responses receive zero correctness and a separate status; their unsupported status is unknown. Reader agreement is never used as the reference answer.

\subsection{Study I: actor accounts and native inputs}
The first $2\times2$ comparison uses R00 (tool log), R10 (tool log plus the saved paragraph, Basis, and Note), R01 (tool log plus native inputs), and R11 (both additions). The native bundle includes the authentic request, recorded input content, channels, chronology, and ID mapping. These are bundled interventions, not isolated estimates for each field. All conditions retain the same original tool log and target. Incident labels, attack assignments, clean authorization references, previous readings, and answer references are withheld.

Q2 is the original primary endpoint because all four conditions provide the relevant literal evidence. The paired comparisons are R10 minus R00 and R11 minus R01. Q1 and Q3 are secondary, with answerability explicitly reported. Luna first reads all 256 packets. Sonnet subsequently reads all the same packets under the same text, schema, references, and scoring, after the Luna results motivated a second-reader comparison. This sequencing is retrospective, not an independently held-out replication of the paper's framing.

\subsection{Study II: binding availability and identifier spelling}
The second $2\times2$ comparison retains the same 64 cases. C00 is a fresh reading of the R10 evidence. C10 adds the complete saved ID-to-record-locator table, without adding user or document text. C01 consistently replaces all actor-local IDs in the account with opaque bijective labels. C11 combines that renaming with the corresponding table. Table entries unrelated to the target are retained in their original order. Renaming preserves tool results, arguments, action order, target literals, output locators, and recorded ID-to-record relations; an integrity check verifies that the immutable tool logs contain no local IDs requiring replacement. A case-specific seed derived from 42 generates opaque labels of the form \code{ref\_} followed by 16 hexadecimal characters.

The common Study II prompt defines the table and its limits. Consequently, C00 is a contemporaneous control under a new prompt, not a byte-identical repeat of R10. We do not interpret their between-study difference as a mapping effect. A table entry pointing to \code{request.user} does not supply the request text: Q1 requires abstention in every C condition. Q2 serves as a negative-control question because its literal evidence is unchanged. The new study's primary comparisons are Q3 grounded correctness for C10 minus C00 and C11 minus C01, separately by reader. This endpoint was fixed before the new outputs and does not replace Study I's Q2 endpoint.

Before the new readings, structural analysis identifies 28 cases requiring the binding table and 36 not requiring it. Table~\ref{tab:schema} specifies the implemented relation and its visibility boundaries. No timestamp, hash, or causal-dependence field is added by a Study II mapping row.

\input{tables/binding_schema}

Within each case and condition, Luna, Sonnet, and the deterministic comparator receive the same evidence packet. This equality holds within a condition; the evidence available across conditions differs according to the specified intervention. The comparator has no access to the private reference and applies literal matching and visible-table joins to the supplied packet. Its Q3 decisions separate supported nonempty sets, empty sets, observed Basis absence, and abstention (Table~\ref{tab:comparator}). Without a table, 18 cases support a Basis-absent finding and 18 an empty set; the remaining 28 require abstention. With a table, 24 support nonempty sets, 22 empty sets, and 18 Basis-absent findings. Thus all 64 grounded decisions are correct in either condition, but they are not 64 identified sources. Q1/Q2 grounded decisions are also correct throughout; Q1 is abstention whenever the genuine request is withheld. For these literal and identifier questions, matching and table joins provide an explicit packet-level validation procedure. The reader comparison tests whether generated findings respect the same evidence boundaries.

\input{tables/comparator}

\subsection{Execution, pairing, and uncertainty}
The readers are \code{gpt-5.6-luna} and \code{claude-sonnet-4-6}. Luna's earlier qualification score was 40/48 under the original strict conditions, below the original acceptance criterion. Earlier human review concerned those qualification labels; the reconstruction references in the present studies are derived mechanically. Both use temperature zero and 2,048 output tokens; Luna uses reasoning effort none and seed 42, while Sonnet uses disabled thinking and has no exposed API seed. Input bounds are 16,384 tokens. Each request is a fresh conversation, with no retries or cross-condition answer sharing. The text and JSON schema are identical across readers, serialized in their native APIs. Study II shuffles case blocks and the eight reader-by-condition requests within each block using seed 42, then executes the frozen order serially. Method and input hashes are sealed before each new pass. Technical/usage failures would stop execution; schema-valid wrong answers remain data and do not trigger retries or sample expansion.

All 1,024 reconstruction assignments were completed, 512 per study, as repeated readings of 64 cases nested in 13 tasks. Pairing retains the case identifier, selected action, and target literal across conditions. Each contrast therefore concerns a change in the supplied evidence for the same recorded action. We average paired correctness changes within each task and then across the 13 task means, so each task contributes equally despite different numbers of applicable cases. Let $t$ index a task, $n_t$ be its number of cases, and $c_0,c_1$ denote the baseline and comparison conditions. Using the question-specific correctness indicator $G$, the reported task-equal contrast is
\begin{equation}
 \widehat\Delta=\frac{1}{13}\sum_{t=1}^{13}
 \frac{1}{n_t}\sum_{i\in t}(G_{ic_1}-G_{ic_0}).
\label{eq:paired}
\end{equation}
In Equation~\ref{eq:paired}, $G_{ic_1}-G_{ic_0}$ is $+1$ when a case changes from incorrect to correct, $-1$ for the reverse change, and zero when correctness is unchanged. The inner average gives each task's change; the outer average gives the overall contrast. Multiplying by 100 expresses the result in percentage points. This differs from pooling all cases, which gives more weight to tasks with more cases. The indicator refers to Q2 for Study I's primary comparison and Q3 for Study II's primary comparison; the questions are not combined.

We resample whole tasks with replacement for 10,000 percentile-bootstrap draws, seed 42. Each bootstrap draw selects 13 tasks with replacement and recomputes the average of their task-level changes. All cases within a selected task stay together, preserving the clustering used by the original analysis. Bootstrap validity depends on how cluster dependence is modeled~\cite{clusterbootstrap}. Given the 13 selected task clusters, we treat the intervals as exploratory descriptions, not confirmatory tests, equivalence bounds, or new acceptance gates. Pooled counts and task-equal changes answer different weighting questions and are reported separately. No cross-question aggregate accuracy is used as a headline outcome.

\subsection{Reproducibility and retained evidence}

Retained artifacts include native requests and responses, evidence views and selection manifests, complete references and visible packets, renaming tables, frozen prompts and schemas, reader outputs, scoring code, run metadata, and usage records. The mapping analysis independently recomputes all completed Q2/Q3 results from visible packets. Failed episodes, inapplicable entries, unresolved outcomes, and reader errors remain auditable. An anonymized artifact package for Studies I and II is available for review\footnote{\url{https://anonymous.4open.science/r/2027_DFRWS_EU-624C}} and traces each reported comparison through the case manifest, visible packet, reference, retained output, and scoring decision.

To verify a reported source, the locator must first identify an eligible earlier external record containing the target literal. A Q3 finding also requires a visible binding from the actor's saved Basis to that record.

%% file: tables/binding_schema.tex
\begin{table}[t]
\centering
\caption{Implemented record-binding contract. IDs and locators are scoped to one saved execution; the case manifest identifies that execution. A mapping row supplies a relation, not record text.}
\label{tab:schema}
\begin{tabularx}{\columnwidth}{@{}lY@{}}
\toprule
Field or structure & Meaning and boundary \\
\midrule
\code{case\_id} & Packet identity; manifest links to the saved run. \\
\code{target.action\_id} & Selected action; ordered tool log defines the earlier-record boundary. \\
\code{stated\_refs} & Saved Basis: null means absent; an empty array means recorded empty. \\
\code{id\_mapping} & Rows contain only \code{id} and \code{locator}; present in C10/C11. \\
Record locator & Whole tool result, one-based list element, or \code{request.user}; no substring offset. \\
\code{native\_inputs} & R01/R11 provide input text, ID, locator, and channel. A user locator alone does not reveal user text. \\
\bottomrule
\end{tabularx}
\end{table}

%% file: tables/comparator.tex
\begin{table}[t]
\centering
\caption{Packet-only comparator, Q3, $N=64$ per condition. Every row has $G=64$ and $U=0$. $D$ includes correct nonempty sets ($+$), empty sets ($\emptyset$), and explicit Basis-absent findings ($M$); $A$ is justified abstention. Grouped conditions have identical counts.}
\label{tab:comparator}
\begin{tabular}{lrrrrr}
\toprule
Condition & $D$ & $A$ & $+$ & $\emptyset$ & $M$ \\
\midrule
R00, R01 & 0 & 64 & 0 & 0 & 0 \\
R10 & 36 & 28 & 0 & 18 & 18 \\
R11 & 64 & 0 & 24 & 22 & 18 \\
C00, C01 & 36 & 28 & 0 & 18 & 18 \\
C10, C11 & 64 & 0 & 24 & 22 & 18 \\
\bottomrule
\end{tabular}
\end{table}

%% file: sections/05-Results.tex
\section{Results}
\subsection{No observed retrieval gain from the actor account in Study I}
Table~\ref{tab:study1} shows every reader/condition result for the three questions. On Study I's primary Q2 endpoint, Luna scores 60/64 with tools alone and 55/64 after adding the actor account; with native inputs, the counts are 64/64 and 63/64. Sonnet scores 64/64, 63/64, 64/64, and 64/64. Task-equal contrasts and exploratory intervals appear in Table~\ref{tab:contrasts}. The counts show no observed retrieval gain, without establishing harm or equivalence. Interpretation is constrained by Sonnet's ceiling and the 13 task clusters.

\input{tables/study1}

With the user request withheld, Luna makes one and nine unsupported Q1 assertions in R00 and R10; Sonnet makes none. Sonnet makes five in R11, so the Luna pattern is not replicated and neither reader dominates all conditions. With the actor account withheld in R00/R01, Luna's 64/64 grounded Q3 scores are abstentions. Sonnet makes six incorrect definite Q3 declarations in R01.

In R10, Luna and Sonnet make 27 and 28 unsupported Q3 assertions, of which 18 and 23 match the complete reference. A post-hoc intersection finds 16 shared reference-matching unsupported answers across seven tasks. These observations motivated Study II, which retains all 64 cases independently of reader correctness. The account bundle increases Luna input tokens by 18.1\% without native inputs and 11.7\% with them. This differs from the original paragraph-only L0/L1 comparison and does not measure human review time.

\subsection{Bindings and the support for recorded citations}
The pre-output mapping-required stratum explains the distinction between evidential and factual gains (Table~\ref{tab:required}). Without bindings, Luna correctly abstains in 7/28 cases and makes 21 unsupported assertions; 12 of those assertions match the complete reference. Sonnet abstains in only 2/28 and makes 26 unsupported assertions, 22 of which match the reference. Sonnet nevertheless retrieves every Q2 literal-location set correctly in C00. The discrepancy concerns the citation-to-record relation despite correct literal retrieval.

\begin{figure*}[!t]
\centering
\includegraphics[width=.94\textwidth]{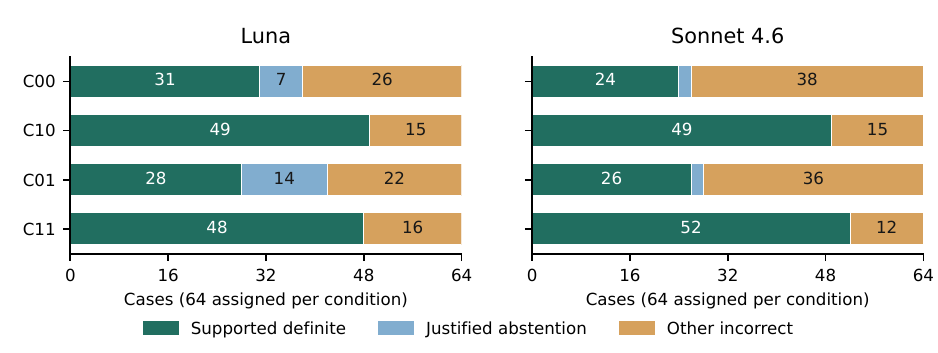}
\caption{Study II Q3 outcomes, retaining all 64 assignments per condition and reader. Grounded correctness is the sum of supported definite correctness and justified abstention, not a synonym for resolved sources. C00/C10 retain original IDs; C01/C11 use opaque IDs. C10 and C11 supply the corresponding complete binding table. The same-packet deterministic comparator is grounded-correct throughout.}
\label{fig:q3}
\end{figure*}

\input{tables/required}
Table~\ref{tab:study2} gives all Study II conditions, and Figure~\ref{fig:q3} separates supported findings from justified abstention. With original identifiers, adding the mapping table changes Q3 grounded counts from 38/64 to 49/64 for Luna and 26/64 to 49/64 for Sonnet. With opaque identifiers, the corresponding counts are 42/64 to 48/64 and 28/64 to 52/64. The task-equal changes are +14.36 and +13.08 percentage points for Luna, and +31.54 and +35.77 for Sonnet. The Luna opaque-ID interval includes zero (Table~\ref{tab:contrasts}); these are not confirmatory significance claims.

\input{tables/study2}
\input{tables/contrasts}

Adding the mapping yields supported definite Q3 answers in 21/28 of these cases for Luna and 24/28 for Sonnet. Across all 64 cases, however, complete-reference agreement changes only from 43 to 49 for Luna and 46 to 49 for Sonnet with original IDs. Some previously correct guesses become supportable because the missing relation is now present. The difference between these measures reflects the newly available support for a finding as well as changes in reader responses; grounded correctness and complete-reference agreement capture distinct outcomes.

The 36 other cases do not show uniform improvement. Their grounded Q3 counts in C00/C10/C01/C11 are 31/28/28/29 for Luna and 24/25/26/27 for Sonnet. Both readers continue to make errors with complete tables. C10 has 15 nongrounded answers for each reader, and C11 has 16 for Luna and 12 for Sonnet. A post-hoc mechanical audit of these 58 erroneous responses (31 distinct cases) identifies 21 unwarranted abstentions, 26 answers containing an extra locator, and six strict-subset omissions. These flags may overlap and do not cover all erroneous responses. They describe response patterns without establishing causes. One extra locator points to the target action result, and none names a nonexistent packet record. Binding preservation therefore does not ensure a correct model reading.

\subsection{Renaming, controls, and shared errors}
Opaque renaming without a table changes Luna's unsupported Q3 assertions in the mapping-required stratum from 21 to 14. Sonnet remains at 26. Every task-equal Q3 renaming-accuracy interval includes zero. Luna's pooled grounded count rises from 38 to 42, yet its task-equal change is $-1.15$ percentage points.

Q3 answers change under renaming in 25/64 cases for Luna and 9/64 for Sonnet without a table, and 16/64 and 7/64 with a table. These changes do not isolate an identifier heuristic because there is only one response per condition and no repeated-identical-condition control for provider nondeterminism. Token sequences also change. Q2 is not exactly invariant despite unchanged literal evidence (Table~\ref{tab:study2}). Its results caution against attributing every Q3 change solely to successful binding use.

Q1 supplies another boundary check. Because the authentic user text is withheld throughout Study II, every justified Q1 answer is abstention. Luna instead makes 9/10/4/10 unsupported assertions in C00/C10/C01/C11; Sonnet makes 0/1/0/0. A mapping to a user-request locator must not be treated as the request itself. All 274 incorrect or unsupported question items among 1,536 Study II question items are retained; these are items, not 274 episodes.

Cross-reader Q3 exact answer agreement is 41/44/28/41 cases in C00/C10/C01/C11. Both readers share a reference-matching but unsupported answer in 11 C00 cases and eight C01 cases, compared with zero in either table condition. These Study II counts differ from the 16-case post-hoc intersection in Study I. Agreement therefore requires checking against the supplied evidence.

%% file: tables/study1.tex
\begin{table}[!htbp]
\centering
\caption{Study I: all counts out of 64. Q1/Q3 G denotes grounded correctness; Q2 is exact complete-set accuracy. Q3 D denotes supported definite correctness and U unsupported assertions. R00: tools; R10: +actor account; R01: +native inputs; R11: both.}
\label{tab:study1}
\setlength{\tabcolsep}{4pt}
\begin{tabular*}{\columnwidth}{@{\extracolsep{\fill}}llrrrrr@{}}
\toprule
Reader & Cond. & Q1 G & Q2 & Q3 G & Q3 D & Q3 U \\
\midrule
Luna & R00 & 63 & 60 & 64 & 0 & 0 \\
Luna & R10 & 55 & 55 & 32 & 31 & 27 \\
Luna & R01 & 63 & 64 & 64 & 0 & 0 \\
Luna & R11 & 62 & 63 & 54 & 54 & 9 \\
Sonnet & R00 & 64 & 64 & 64 & 0 & 0 \\
Sonnet & R10 & 64 & 63 & 24 & 21 & 28 \\
Sonnet & R01 & 64 & 64 & 58 & 0 & 6 \\
Sonnet & R11 & 59 & 64 & 50 & 50 & 7 \\
\bottomrule
\end{tabular*}
\end{table}

%% file: tables/required.tex
\begin{table}[!htbp]
\centering
\caption{Q3 in the 28 cases requiring a mapping. A is justified abstention; D is supported definite correctness; U is unsupported assertion; W\,$\cap$\,U counts unsupported answers matching the complete reference. All 28 assignments per cell remain included.}
\label{tab:required}
\setlength{\tabcolsep}{4pt}
\begin{tabular*}{\columnwidth}{@{\extracolsep{\fill}}llrrrrr@{}}
\toprule
Reader & Cond. & N & A & D & U & W $\cap$ U \\
\midrule
Luna & C00 & 28 & 7 & 0 & 21 & 12 \\
Luna & C10 & 28 & 0 & 21 & 5 & 0 \\
Luna & C01 & 28 & 14 & 0 & 14 & 11 \\
Luna & C11 & 28 & 0 & 19 & 5 & 0 \\
Sonnet & C00 & 28 & 2 & 0 & 26 & 22 \\
Sonnet & C10 & 28 & 0 & 24 & 4 & 0 \\
Sonnet & C01 & 28 & 2 & 0 & 26 & 20 \\
Sonnet & C11 & 28 & 0 & 25 & 3 & 0 \\
\bottomrule
\end{tabular*}
\end{table}

%% file: tables/study2.tex
\begin{table}[!htbp]
\centering
\caption{Study II: all counts out of 64. G, D, and U follow Table~\ref{tab:study1}; W is complete-reference agreement, which can include unsupported guesses. C00/C10: original IDs; C01/C11: opaque IDs; C10/C11 add the matching table. Q1 requires abstention throughout.}
\label{tab:study2}
\setlength{\tabcolsep}{4pt}
\begin{tabular*}{\columnwidth}{@{\extracolsep{\fill}}llrrrrrr@{}}
\toprule
Reader & Cond. & Q1 G & Q2 & Q3 G & Q3 D & Q3 U & Q3 W \\
\midrule
Luna & C00 & 55 & 49 & 38 & 31 & 21 & 43 \\
Luna & C10 & 54 & 52 & 49 & 49 & 9 & 49 \\
Luna & C01 & 60 & 48 & 42 & 28 & 15 & 39 \\
Luna & C11 & 54 & 50 & 48 & 48 & 6 & 48 \\
Sonnet & C00 & 64 & 64 & 26 & 24 & 28 & 46 \\
Sonnet & C10 & 63 & 62 & 49 & 49 & 9 & 49 \\
Sonnet & C01 & 64 & 64 & 28 & 26 & 28 & 46 \\
Sonnet & C11 & 64 & 63 & 52 & 52 & 7 & 52 \\
\bottomrule
\end{tabular*}
\end{table}

%% file: tables/contrasts.tex
\begin{table*}[!t]
\centering
\caption{Primary paired contrasts for each study, in percentage points. Tasks receive equal weight (13 clusters); 95\% percentile intervals use 10,000 task-bootstrap draws. These exploratory intervals do not establish general harm, equivalence, or confirmatory significance.}
\label{tab:contrasts}
\setlength{\tabcolsep}{4pt}
\begin{tabular*}{\textwidth}{@{\extracolsep{\fill}}lllrr@{}}
\toprule
Study / endpoint & Reader & Contrast & Change (pp) & Interval (pp) \\
\midrule
I: Q2 & Luna & R10 $-$ R00 & -7.18 & [-20.51, +3.85] \\
I: Q2 & Luna & R11 $-$ R01 & -1.28 & [-3.85, +0.00] \\
I: Q2 & Sonnet & R10 $-$ R00 & -1.54 & [-4.62, +0.00] \\
I: Q2 & Sonnet & R11 $-$ R01 & +0.00 & [+0.00, +0.00] \\
II: Q3 & Luna & C10 $-$ C00 & +14.36 & [+0.26, +30.00] \\
II: Q3 & Luna & C11 $-$ C01 & +13.08 & [-4.62, +32.56] \\
II: Q3 & Sonnet & C10 $-$ C00 & +31.54 & [+14.87, +51.03] \\
II: Q3 & Sonnet & C11 $-$ C01 & +35.77 & [+15.13, +57.18] \\
\bottomrule
\end{tabular*}
\end{table*}

%% file: sections/06-Discussion.tex
\section{Discussion}
\subsection{Preservation and evidence export}
Retrieving a value and establishing which record a saved citation denotes require different evidence. Study II supplies the ID-to-record relation without adding document or user text. For the tested questions, this restores a directly checkable link while leaving the underlying tool content and actions unchanged.

Evidence collection and export should retain the native tool calls and results, their execution order, the authentic request and channel identity, the public actor account, and execution-scoped input-ID bindings. Table~\ref{tab:schema} describes the binding relation implemented here. Generated accounts should remain identifiable as generated artifacts, and derived views should identify withheld fields. A locator alone does not supply the content of the record it names.

The experiment removes bindings from derived evidence packets rather than measuring their loss in deployed systems. Its preservation implication concerns retaining the relationships needed to check a finding as records are transformed or shared. Hashes and retained originals support alteration detection within the instrumented collection. Source authenticity and operational chain of custody require separate validation.

\subsection{Reporting the supported relation}
The illustrative case \code{rcase-001} separates three reportable questions (Table~\ref{tab:reporting}). Both packets establish that the recipient appears in the earlier bill result. Only the packet with the binding table establishes that the saved Basis refers to that record. The finding describes the recorded citation relationship. It does not establish why the actor selected the recipient.

\begin{table}[!htbp]
\centering
\caption{Findings supported by the C00/C10 packets for \code{rcase-001}. These summarize the existing illustrative case under the declared evidence contract.}
\label{tab:reporting}
\begin{tabularx}{\columnwidth}{@{}lY@{}}
\toprule
Question & Supported finding \\
\midrule
Q1: Request & Unresolved in both packets: authentic request text is withheld. \\
Q2: Occurrence & The recipient appears in \code{act-01.result} before the transfer in both packets. \\
Q3: Citation & Unresolved in C00; the explicit binding in C10 identifies the bill result as the saved Basis reference. \\
\bottomrule
\end{tabularx}
\end{table}

Reporting should also distinguish an empty relation, an observed absent Basis field, and insufficient supplied evidence. An empty set reports the result of the defined record query. A missing field reports the state of the saved account. Neither warrants extending the finding to unobserved actor behavior.

\subsection{Validation of assisted findings}
Accurate literal retrieval, complete-reference agreement, and cross-reader agreement each coexist with unsupported citation findings. Evaluation should therefore report supported definite findings and justified abstention separately. Adding evidence can make a previously correct guess supportable without producing a comparable increase in complete-reference agreement.

For these literal and identifier questions, the packet-only deterministic comparator answered or abstained correctly throughout. It supplies a validation check for relationships that can be established through matching and table joins. The readers still produce 12--16 nongrounded Q3 answers with complete tables, so preserving the required relation and reading it correctly are separate requirements. Findings that depend on an unavailable binding should retain an unresolved status in the assisted output.

For reconstruction review, a generated finding can be retained together with the case and target-action identifiers, the supplied packet, and the locator set it asserts. Verification can then establish literal occurrence in earlier records before checking any execution-local citation binding. This makes a citation claim inspectable as a relation between preserved records. It also keeps the actor's generated account distinct from the independent check of that account. Requirement-based testing of search functions~\cite{guovalidation} provides a precedent for validating these operations separately.

A review output should preserve the distinction between supported nonempty relations, supported empty relations, observed Basis absence, and insufficient supplied evidence. These outcomes have different consequences for reporting. An empty intersection answers the defined query, whereas an unresolved binding leaves a relation unestablished even when the corresponding source is plausible. Keeping the status beside the cited locator allows a later reviewer to see whether a definite finding was justified by the packet, rather than reconstructing that distinction from a fluent narrative.

Support should also be recorded against the evidence version used for the finding. In \code{rcase-001}, supplying the table changes the support for Sonnet's unchanged answer. A later packet can therefore justify a relation that an earlier packet could not establish. Recording which content and bindings were supplied makes that change inspectable without treating the earlier guess as grounded. This documentation perspective is consistent with work on reliability validation of forensic interpretation~\cite{nordvikvalidation}. It locates the reported finding within a reproducible evidence check.

One potential application is reviewing a reduced agent trace after an automated transfer. If the native execution retained the required citation bindings, an exporter can include them with the records used to verify a finding. If a required binding was never recorded, this audit leaves the relation unresolved. This use depends on preserving execution-scoped bindings during collection, before producing an assisted report.

\subsection{Limitations}
The audit verifies literal occurrence, temporal eligibility, and recorded identifier relations. Its findings describe what the supplied packet supports. A saved citation records the actor's stated reference and does not establish semantic authorization, harmful intent, or internal causal reliance.

The corpus contains 64 applicable cases from 13 tasks in one simulated Banking environment, with 32 structurally inapplicable inventory entries retained in the selection manifest. The derived evidence conditions do not estimate the prevalence of missing bindings in operational systems. No investigator evaluation was conducted, so the results do not quantify benefits to casework.

The studies were developed retrospectively, with inputs and analyses fixed before each pass. Sonnet was selected after Luna's results. Repeated readings reuse the same cases, and task-bootstrap intervals are exploratory. The comparison describes two specific reader configurations, with Sonnet sharing the Claude family with the Haiku actor in 17 cases. One response per condition and hosted-model nondeterminism limit the attribution of individual output changes, particularly under identifier renaming.

%% file: sections/07-Conclusion.tex
\section{Conclusion}
Forensic review of a saved agent action requires distinguishing a value recorded in the evidence from a supported link to the actor's stated citation. In this corpus, exact literal retrieval and agreement with the complete reference coexist with unsupported citation findings, including answers shared by both readers. In C00, Sonnet retrieves every literal-location set but makes 26 unsupported citation-source assertions in the 28 mapping-required cases. Twenty-two match the complete reference. This contrast shows why validation must retain the support status alongside the answer when checking whether the supplied records establish the reported relation. Explicit ID-to-record bindings make the relation checkable and improve grounded reconstruction for the measured readers, although complete tables do not eliminate their errors. Identifier renaming alone provides no consistent remedy.

For the tested questions, preservation should retain record content, execution order, channel identity, and the execution-scoped bindings needed to verify a reported relation. Generated accounts should remain distinguishable from the records they reference. The packet-only deterministic comparator provides a check for the literal and identifier relations examined here. Reporting supported definite findings, observed empty or missing states, and unresolved relations separately prevents agreement with a complete reference from being treated as evidence of support in the supplied packet.